# Is there a density-gradient expansion for the electronic exchange energy?

John P. Perdew[1], Timo Lebeda[2]. Rohan Maniar[1], Chandra Shahi[1], Adrienn Ruzsinszky[1], and Jianwei Sun[1]

[1] Department of Physics and Engineering Physics, Tulane University, New Orleans., LA, U.S.A. 70118

[2] Yusuf Hamied Department of Chemistry, Cambridge University, Cambridge, U.K.

Abstract: An old but recurring question in ground-state density functional theory, relevant to the construction of density functional approximations, is the existence of a second-order gradient expansion for the exchange energy that is exact for densities that vary slowly over space. Here we will argue that there is such an expansion, with a gradient coefficient 10/7 of Sham's, as suggested by theoretical/mathematical derivations and by numerical calculations on model slowly-varying metallic densities, although some standard derivations do not quite capture the correct gradient coefficient. We also discuss long-range anomalies in the gradient expansions for the exchange and correlation holes around an electron. It is plausible that, as suggested earlier and recently, standard derivations are still correct for the coefficient of the second-order gradient expansion for exchange and correlation together, in which case the correct gradient coefficient for correlation is $C_c = C_{xc} - C_x$. We propose possible numerical evaluations of $C_{xc}$ and $C_x$ that are free from long-range anomalies, and discuss implications for density functional approximations. Axel Becke's 1988 generalized gradient approximation (the B88 GGA) for the exchange energy brought density functional theory into chemistry, with a gradient coefficient for exchange that is optimal at the GGA level for atoms and molecules but too large for solids. Meta-GGAs seem to work as well for molecules as for solids, but may not be perfected yet. The article concludes with a few memories that the first author has of discussions with Axel Becke.

1 Short history of the gradient expansion for the exchange energy

Modern density functional theory began in 1964-65 with the foundational work of Hohenberg and Kohn[1] and Kohn and Sham[2]. The first gradient expansion and generalized gradient approximation were made by Ma and Brueckner[3] in 1968, not for exchange but for the more complicated correlation energy, to complement Hartree-Fock exchange. In 1969, Herman, Van Dyke and Ortenburger[4] proposed the second-order density-gradient expansion for the exact exchange energy of a spin-unpolarized electron density,

$$E_x[n] = \int d^3r\,(A_x n^{4/3} + C_x|\nabla n|^2/n^{\frac{4}{3}}). \qquad (1)$$

Here $n(\boldsymbol{r})$ is the electron density at position $\boldsymbol{r}$, $A_x$ = -(3/4$\pi$)$(3\pi^2)^{\frac{1}{3}} = -0.73856$ (in atomic units where $e^2 = 1$), and $C_x$ was a negative gradient coefficient fitted to the known exchange energies of atoms. The first term of Eq. (1) is the local density approximation for exchange, known to Kohn and Sham, and the second order gradient term was constructed from dimensional considerations. (We now know that the form of Eq. (1) is dictated by the uniform density scaling property[5] of the exact exchange energy. The exact exchange energy is the Hartree-Fock exchange energy evaluated with exact Kohn-Sham orbitals (implicit functionals of the density) in place of Hartree-Fock orbitals.)

In 1971, Sham[6] derived Eq. (1) with $C_x = C_x^{Sham} = -7/[432\pi(3\pi^2)^{\frac{1}{3}}] = -0.001667$, a value much smaller in magnitude than the value fitted[4] to atoms. Like Ma and Brueckner and many others, Sham started by applying a sinusoidal static density variation of small amplitude $A_q$ with wavevector $q$ to a system of uniform density $\bar{n}$. The resulting change in the exchange-correlation energy per unit volume is, to lowest order in $A_q$, $K_{xc}(q)A_q^2$. The function $K_{xc}(q)$ is found by many-body perturbation theory within or beyond the random phase approximation as an integral over internal wavevectors $k$, and expanded as $a_{xc}$+$b_{xc}q^2 + \cdots$. When $q$ is small, the density is slowly-varying, the density gradient expansion applies (although it also applies to slowly-varying densities of large amplitude), and the coefficients of the small-$q$ expansion determine the gradient coefficient $C_{xc}\,(\overline{n}) = C_x + C_c(\overline{n})$. The Ma-Brueckner[3] calculation of $C_c^{MB}(\infty) = 0.004235$ was extended to finite $\bar{n}$ within the random phase approximation (RPA) by Geldart and Rasolt[7] 1976; this coefficient decreases as $\bar{n}$ decreases. All these calculations regularize the bare long-range electron-electron Coulomb repulsion $1/R$ (with Fourier transform $4\pi/k^2$) to the finite range $e^{-\kappa R}/R$ (with Fourier transform $4\pi/(k^2+\kappa^2)$, and take the limit $\kappa \to 0$ at the end of the calculation. The internal wavevector $k$ in all these calculations is the momentum argument of a Green's function. A different wavevector analysis, with a different internal wavevector $k$ of density fluctuations, was made by Langreth and Perdew[8] in 1980. It gave the same integrated $C_x$ that Sham found, and an integrated $C_c(\overline{n})$ similar to that of Geldart and Rasolt. The gradient expansion coefficients seemed to be well determined. (Both wavevector decompositions, for exchange and for correlation, included integrable distributions like the Dirac delta function $\delta\left(k - 2\overline{k_F}\right)$, where $2\overline{k_F}$ is the diameter of the spherical Fermi surface, and $\overline{k_F} = (3\pi^2\bar{n})^{1/3}$.)

But a larger exchange-only gradient coefficient, $(10/7)C_x^{Sham} = -0.002381$, was derived and evaluated earlier by Geldart and Taylor[9] 1970 and later by Kleinman and Lee[10] in 1988, Vosko and collaborators[11,12] in 1990, and Svendsen and von Barth[13] (who also derived fourth-order gradient coefficients) in 1996. The Kleinman-Lee argument was appealing: The inverse lengths $q$ and κ must both tend to zero, but the order in which these limits are taken matters, because $\frac{4\pi}{k^2}$ is singular at $k$=0. Sham took the $q \to 0$ limit first, but physics requires taking the $\kappa \to 0$ limit first, since the bare Coulomb potential needed in the exchange energy is always long-ranged.

In 1996, Springer, Svendsen and von Barth[14] created slowly-varying model densities for metallic solids, and evaluated their exact exchange energies, confirming that their gradient expansion (with $C_x = (10/7)C_x^{Sham}$) gave the right correction to the local spin density approximation for the exchange energy. Their model densities were not necessarily at the slowly-varying limit, but they were more slowly-varying than any realistic electron density and thus directly relevant to the construction of density functional approximations for realistic electron densities.

The separation of the exchange-correlation energy into separate exchange and correlation energies is needed for the construction of density functional approximations from the known mathematical properties of the impractical exact density functional for the exchange-correlation energy. Many of those properties are for the exchange energy alone. The first true GGA to come from the Perdew group[15], and the most-widely used one[16] did not employ either $C_x^{Sham}$ or $(10/7)C_x^{Sham}$, but $(10/7)C_x^{Sham}$ was used in the PW91 GGA[17], and more naturally in the PBEsol[18] GGA for solids in 2008. $(\frac{10}{7})C_x^{Sham}$ was also used in meta-GGAs including TPSS[19] (2003), SCAN[20] (2015), r2SCAN[21] (2020), and LAK[22] (2024).

## 2 Position-space insights into the gradient expansion for exchange

For many of us, position space is more intuitive than wavevector space. Under the assumption of adiabatic connection[23,24] between a ground state wavefunction for real electrons and one of non-interacting electrons of the same electron density, the exact exchange-correlation energy can be written as

$$E_{xc}[n] = \frac{1}{2}\int d^3r \int d^3r' \, n(\boldsymbol{r}) n_{xc}(\boldsymbol{r},\boldsymbol{r}')/|\boldsymbol{r}'-\boldsymbol{r}|, \tag{2}$$

or by an equivalent expression[23] in wavevector space. Here $n_{xc}(\boldsymbol{r},\boldsymbol{r}') = n_x(\boldsymbol{r},\boldsymbol{r}') + n_c(\boldsymbol{r},\boldsymbol{r}')$ is the coupling-constant-averaged density at $\boldsymbol{r}'$ of the exchange-correlation hole surrounding an electron at $\boldsymbol{r}$. Eq. (2) tells us that[24] this energy depends on the hole density only through its system- and spherical-average. The exchange contribution can be computed from the non-interacting one-particle density matrix $\rho(\boldsymbol{r},\boldsymbol{r}')$ constructed from the occupied exact Kohn-Sham orbitals.

Among the earliest known exact constraints in density functional theory are those on the exchange and correlation holes[23,24]. The exact exchange hole density is negative, integrates to -1, and takes the value $-n(\boldsymbol{r})/2$ when $\boldsymbol{r}' \to \boldsymbol{r}$ for a spin-unpolarized density. The exact correlation hole integrates to 0 and has a Coulomb cusp[25] at $\boldsymbol{r}' \to \boldsymbol{r}$. All these conditions are satisfied by the local density approximation (LDA), in which the hole is that of a possible system (the uniform electron gas), and this fact explains[23.24] why LDA works as well as it does for real spin-unpolarized systems. But, in the second order gradient expansion, the correlation hole density does not integrate to zero[8,23], and that is why the un-generalized gradient expansion for correlation is very seriously in error for real systems, strongly requiring a generalized gradient approximation[3,26] that only recovers the gradient expansion in the limit of slowly-varying density.

In 1981, Gross and Dreizler[27] derived a gradient expansion of the exchange energy in position space. Unlike other derivations of the gradient coefficient, they did not assume a density variation of small amplitude. Their exchange energy density has contributions proportional to $|\nabla n|^2$ and to $\nabla^2 n$, and the latter contribution has a divergent integral in its coefficient. Integrating by parts over the system eliminates that divergence, leaving only $|\nabla n|^2$ with at worst integrals that can be regularized by the factor exp($-\kappa R$) with $\kappa \rightarrow 0$. With this regularization, Gross and Dreizler recovered Sham's gradient coefficient $C_x^{Sham}$.

Perdew[28] concluded that the gradient expansion of the exchange hole density improves upon LDA at positions $\boldsymbol{r}'$ close to the electron at $\boldsymbol{r}$, but fails badly at positions far away. That is why the un-generalized gradient expansions fail badly for real systems. Cutting off the spurious long-range parts of the gradient expansion for the exchange hole density to satisfy the known exact constraints above led to two exchange-energy functionals[28,15] that were more accurate than LDA for real systems. Only the second one[15] was a true GGA, depending only on $|\nabla n|^2$.

It is not widely known that the much-used PBE GGA has two different constructions[16,29] that lead to nearly the same functional. One is the satisfaction of selected exact constraints on the density functional for the exchange-correlation energy, in Ref. 16. The other[29] applies sharp separate long-range position-space cutoffs to the gradient expansions of the exchange and correlation holes to satisfy the known hole constraints above. The gradient expansion of the exchange hole was the one used in Ref. 15. The gradient expansion of the correlation hole was constructed by Fourier transformation of the Langreth-Mehl[26] long-range part, with a short-range correction to recover the Ma-Brueckner[3] gradient coefficient $C_c^{MB}(\infty)$. Note that the correlation hole in position space can be recovered from its Langreth-Mehl wavevector analysis because the wavevector $k$ of the latter is the wavevector of a density fluctuation. In the PBE GGA, the gradient expansions for exchange and correlation are chosen to cancel each other. This cancellation was consistent with but not required by the position-space cutoffs of the gradient expansions for the exchange and correlation holes. The cancellation built into PBE[16] and PBEsol[18] was suggested by the LDA-like linear response function of the uniform electron gas found in diffusion quantum Monte Carlo[30] calculations at valence-electron densities.

It is not at the GGA but only at the meta-GGA level that a functional seems flexible enough to satisfy many exact constraints and to work well for both molecules and solids. The second-order gradient expansion of the exact exchange-correlation energy is expected to be accurate for regions of space where the GGA ingredients

$$s = \frac{|\nabla n|}{2k_F n} \tag{3}$$

and (for correlation)

$$t = \frac{|\nabla n|}{2k_{TF} n} \tag{4}$$

are much less than 1, and the meta-GGA ingredient

$$\alpha = (\tau - \tau_W)/\tau_{unif} \geq 0 \tag{5}$$

is close to 1. Here $\tau$ is the positive non-interacting or orbital kinetic energy density, and $\tau_W$ and $\tau_{unif}$ are the one-electron von Weizsäcker[31] and uniform density limits of $\tau$. The local Fermi wavevector and the Thomas-Fermi screening wavevector are respectively

$$k_F = (3\pi^2 n)^{1/3} \tag{6}$$

and

$$k_{TF} = (4k_F/\pi)^{1/2}. \tag{7}$$

The discussion in sections 1 and 2 reveals long-standing uncertainties about the gradient coefficients for the exchange and correlation energies. Resolving these uncertainties might lead to better-founded and more accurate meta-GGAs.

## 3 Recurring questions and how they might be answered

In 1987, Langreth and Vosko[32] stated that, despite the controversy over the gradient coefficient for exchange, the sum of the regularized coefficients for exchange and correlation together should cancel out of the leading Feynman diagrams for $C_{xc}$ at high density. In 2011, the treatise by Engel and Dreizler[33] suggested that (10/7)$C_x^{Sham}$ is right for exchange, and the Geldart-Rasolt 1976 $C_{xc}$ (whose exchange-like part is $C_x^{Sham}$) is right for exchange+correlation.

In 2026, Benites, Rosado, and Manousakis[34] revisited calculations of the gradient coefficients for exchange and correlation. They used a regularizer *exp(-* $\kappa\beta(k_F)R)$ with $\kappa \to 0$. With $\beta(k_F)$ independent of $k_F$, they recovered the Sham $C_x^{Sham}$ and a Geldart-Rasolt or Ma-Brueckner $C_c^{BRM}(\infty)$ in the high-density limit. But otherwise the exchange and correlation coefficients were changed in an equal but opposite way by a given $\beta(k_F)$, so that their sum was always the same. From these results, they drew two conclusions: (1) There is no separate gradient expansion for exchange alone or correlation alone. (2) There is a gradient expansion for exchange and correlation together, with their gradient coefficient in the high-density limit.

Their second conclusion may be correct. Their first conclusion is consistent with their results, but seems inconsistent with other evidence:

(i) The analytical derivation of the exchange coefficient (10/7) $C_x^{Sham}$ by Engel and Vosko[12] does not require a regularizer and the only assumption made in Ref. EV's proof was rigorously proven to be true in later work[35].

(ii) The numerical tests of Springer, Svendsen, and von Barth[16] 1996 suggest that there is a gradient expansion for exchange, with gradient coefficient (10/7)$C_x^{Sham}$. Furthermore, the $k \to 0$ limit of the one-dimensional integrand for the Sham exchange coefficient is[26] proportional to $-k/k_F$, while the $k \to 0$ limit of the corresponding integrand for the Ma-Brueckner correlation coefficient is[26] a high peak approximately proportional to $\exp(-\frac{ck}{k_{TF}})$, where $k_{TF}$ is the Thomas-Fermi screening length (defined in Eq. 7), suggesting that the former integrand is exchange and the latter is truly correlation, since only correlation can have the length scale $k_{TF}^{-1}$. (The Langreth-Mehl[26] GGA was constructed by cutting off all contributions to the peak from $k$ less than a sharp wavevector-space cutoff.) Moreover, the long-range anomalies of the gradient expansion for the system- and spherically averaged exchange and correlation hole densities seem very different and non-cancelling, so it is hard to see from this perspective why they should be more correct together than they are separately. The gradient contributions to the hole densities do seem to be correct at short range, and in that sense can be used with separate long-range cutoffs[15] to produce a reasonable GGA[29] for exchange and correlation.

(iii) The exchange energies of neutral atoms of large atomic number $Z$ also support the existence of a gradient expansion for the exchange energy with gradient coefficient $(10/7)C_x^{Sham}$. For large $Z$, LDA exchange gives the leading contribution of order $Z^{5/3}$, followed by known correction terms of order $Z$ and $Z^{2/3}$, to which the SCAN meta-GGA[20] (with gradient coefficient (10/7) $C_x^{Sham}$) was fitted. The term of order $Z$ comes partly from a deep-core region close to the nucleus where the density is not slowly-varying (because $\alpha \ll 1$), and partly from a surrounding region of high but slowly-varying density (with $\alpha \approx 1$ and $s \ll 1$) described by the second-order gradient expansion for exchange. The accurate asymptotics of SCAN exchange, and its small relative errors (less than 0.5%) for Ne, Ar, Kr, and Xe, are most naturally explained by the correctness of SCAN exchange in both regions. While the fitting might disguise some error in the gradient coefficient, it is hard to see how it could disguise the non-existence of a gradient expansion for exchange. (Recent work[36] on the large-$Z$ expansion for the exchange energy finds that our Eq. (1) and numerical fits to exact-exchange calculations both lead to the more correct form $a_x Z^{\frac{5}{3}} + b_x Z ln Z + c_x Z + \cdots$, where the $Z ln Z$ arises from the deep core density, which with $\alpha \ll 1$ is not slowly varying even when $Z \rightarrow \infty$.)

There may be a way to answer these questions, following but improving the model solids of Springer, Svendsen, and von Barth[14] 1996: Imagine a fixed, close-packed lattice of local pseudopotentials embedded in a uniform gas of density $\bar{n}$, characterized by a Fermi wavevector and a Thomas-Fermi screening wavevector, both of which increase without bound as $\bar{n}$ increases. The ratio of a fixed non-zero reciprocal lattice vector $G$ to either characteristic wavevector of the uniform gas will then decrease toward zero as $\bar{n}$ increases, and the slowly-varying limit will be approached. (Ref. 13 employed an fcc lattice with $G/(2k_F) = 0.77$ and 0.89, but substantially smaller ratios can be achieved by increasing $\bar{n}$.) Compute the exact exchange energy, and the RPA approximation to the correlation energy (now available in several solid-state computer codes). In this way, determine the gradient coefficients from exchange and correlation holes that have no anomalous long-range behavior and become increasing localized within the fixed lattice unit cell as $\bar{n}$ increases.

For exchange alone, there might be an even simpler numerical test, proposed but not implemented in Ref. 37: Start with a smooth two-electron density, then apply the Thomas-Fermi scaling[37] to scale it into a density of the same shape with an ever-larger even-integer number of electrons. Use an existing Kohn-Sham inversion to find the exact Kohn-Sham one-electron effective potential and exact Kohn-Sham orbitals, and evaluate the exact exchange energy from the orbitals. Then approach the limit of large particle number in which the density becomes slowly-varying for exchange (but not for correlation). But, unlike the model solids approach, this scaling approach might be defeated by contributions[38] from the low-density "edge" region in which the density variation is rapid.

A 1982 numerical study[3] which seemed to support $C_x^{Sham}$, and which we expect would have no edge contribution in its slowly-varying limit, is discussed in Appendix A, where we show that it supports the existence of a gradient expansion for exchange but does not get close enough to the slowly-varying limit, or have accurate enough reference values, to distinguish between $C_x^{Sham}$ and (10/7) $C_x^{Sham}$.

We have two concerns regarding Ref. 34 that may explain their different conclusion. First, the use of density-dependent regularizers is questionable from a physical point of view: In the work of Ref. 34, Eq. (63) shows a contribution to $C_x$ from $\partial/\partial k_F \beta(k_F)$ that is not present in $C_{xc}$ according to Eq. (81). But, to make the position-space electron-electron interaction depend on $\boldsymbol{r}$ and $\boldsymbol{r'}$ only through $|\boldsymbol{r}' - \boldsymbol{r}|$, and thus to make its Fourier transform depend only on the corresponding variable $k$, $\beta(k_F)$ must be a true constant, independent of $\boldsymbol{r}$ and $\boldsymbol{r'}$. So the appearance of its derivative is unexpected. If we think of $\kappa' = \kappa\, \beta(k_F)$ as just another positive constant like κ, then the limits as each constant tends to zero should be the same. Second, Ref. 34 physically overinterprets their mathematical result. The fact that the singularity of the Coulomb interaction is non-analytical and sensitive to different regularization schemes is well-established[13]. However, this does *not* imply the non-existence of a gradient expansion of the exchange energy. Instead, this underscores that care must be taken to respect the correct physics when choosing regularizers and taking limits. As pointed out by Kleinman and Lee[10], the physics of the (unscreened) exchange energy suggests taking the long-wavelength limit for the bare (unregularized) Coulomb interaction first. Although the derivations by Kleinman and Lee[10] and Svendsen and von Barth[13] that take the limit of vanishing regularization first still contain integrals that must be regularized in some way, the analytical derivation by Engel and Vosko[12] yields the same result without requiring a regularization[35], thus suggesting that the regularizations by Kleinman and Lee[10] and Svendsen and von Barth[13] respect the underlying physics. Ref. 35 reaches the unambiguous mathematical conclusion that $C_x$=(10/7) $C_x^{Sham}$.

## 4 Possible implications for meta-generalized gradient approximations

Because generalized gradient approximations (GGAs) improperly treat all regions of small $s$ and $t$ as regions of slowly-varying density, having the right gradient coefficients for exchange and correlation is not so important for their accuracy. Meta-GGAs, which revert to the gradient expansions only in regions where in addition $\alpha$ is close to 1, should benefit more from having the right gradient coefficients. Even the conditions $\alpha \approx 1, s \ll 1,$ and $t \ll 1$ are only necessary and not sufficient to identify a region of slowly-varying density, but they are in practice much more effective for this identification than the conditions $s \ll 1$ and $t \ll 1$ alone.

In 2003, Perdew and Kurth[39] wrote a review article which accepted (10/7)$C_x^{Sham}$ as the gradient coefficient for exchange, but concurred with Langreth and Vosko[32] 1987 that the Geldart-Rasolt[7] 1976 gradient coefficient (whose exchange part is $C_x^{Sham}$) is correct for exchange and correlation together. While Benites, Rosado, and Manousakis[34] 2026 reject all gradient coefficients for exchange, they have given the best explicit argument (paraphrased here) for accepting the Geldart-Rasolt $C_{xc}$,: It is derived in the RPA, which employs a physically-correct screened effective electron-electron interaction whose range in the high-density limit is proportional to $k_{TF}^{-1}$. Since $q$ can easily be much less than $k_{TF}$, there is no order-of-limits problem for exchange and correlation together in RPA-like calculations. The long-range anomalies of the gradient expansion for the exchange-correlation hole are still there, and still raise some doubt.

Perdew and Kurth[39] 2003 also discussed the PKZB[40] and TPSS[19] meta-GGAs. In a striking disconnect, these meta-GGAs and their successors, the SCAN[20] and r²SCAN[21] meta-GGAs all use $(10/7)C_x^{Sham}$ for exchange but the Ma-Brueckner[3] 1968 or Geldart-Rasolt[7] 1976 (GR) gradient coefficient for correlation. To our knowledge, the more plausible alternative

$$C_c(n) = C_{xc}^{GR}(n) - (10/7)C_x^{Sham} > C_c^{GR}(n) \tag{8}$$

has so far been employed only in the LAK[22] meta-GGA. Perhaps that is why the relative errors of SCAN and r²SCAN for the exchange energies of the rare-gas atoms are so small (less than 0.5%), while they are much larger for the correlation energies[20]. Since a recent attempt[42] to restore the neglected Rasolt[43,44] 1977 contribution from the gradient of the relative spin polarization was not based on Eq. (8) and since LAK considered only the spin-unpolarized limit of Eq. (8), we plan to test the spin-polarization generalization of Eq. 8 in a possible revision of r²SCAN and LAK. (Recent work[45] suggests that exactness for all uniform and one-electron densities are the most important exact constraints on the density functional for the exchange-correlation energy.)

After this section was written, Kieron Burke called our attention to Shegelski, Geldart, Glasser and Nelson[46] 1994. Our interpretation of a statement in this article is that, for the undamped/unregularized Coulomb interaction, the right gradient expansion for exchange has the coefficient $(10/7)C_x^{Sham}$, and the right expansion for exchange+correlation has the coefficient $C_{xc}^{GR}(n)$, at least in the high-density limit, consistent with Eq. (8).

## 5 Axel Becke and his 1988 GGA

Some of the controversies discussed here were already raging in 1988, when Axel Becke created his B88[47] GGA. Axel was not concerned about a slowly-varying limit that is not important to a GGA in atoms and molecules, so he fitted his B88 GGA to the Hartree-Fock energies of the rare-gas atoms, with a gradient coefficient for exchange slightly bigger in magnitude than that for PBE, and much larger in magnitude than any candidate for the slowly-varying limit. As pointed out in Ref. 33, this makes B88 almost exact for the leading correction to LDA in the large-$Z$ expansion for the exchange energies for neutral atoms of atomic number $Z$. B88 did recover LDA in the slowly-varying limit, and Ref. 36 also points out that LDA becomes relatively exact for the exchange energies of neutral atoms in the large-$Z$ limit, showing that (contrary to expectations expressed in section I of Ref. 7) LDA is a good starting point to design density functional approximations for atoms and molecules, and not just for solids.)

The first author of this article never worked with Axel Becke, but had the opportunity to talk with him at chemistry conferences, starting in the early 1990's when B88 caught the interest of many theoretical chemists. Here are a few recollections:

How Axel created the hybrid functionals: He knew that Hartree-Fock theory, with exact exchange and no correlation, underbinds atoms in molecules, while the LDA and GGA functionals, with approximated exchange and correlation, overbind. This led to the idea of a linear mixing of Hartree-Fock exchange with LDA or GGA exchange, plus full LDA or GGA correlation, to predict realistic bonding. Later it occurred to him that the adiabatic connection integral for the exact exchange-correlation energy also suggests this mixing[49], because its integrand is exact exchange at the lower limit of a non-interacting wavefunction and is the non-Hartree part of the expectation value of the electron-electron interaction at the upper limit of the real interacting wavefunction, along a path of fixed density.

Theorizing as daydreaming: Axel said that his best and most creative ideas were the product of "daydreaming" or free-roaming intuition.

From alpha to omega, and back again: His early density functionals had few parameters and were shaped more by physical insight. At mid-career, he developed functionals with many empirical parameters fitted to molecular energy differences. But near the end of his career he returned to few-parameter functionals, in which each parameter could be guessed and was only fitted as a sort of final refinement.

Many university professors don't want to retire, but Axel looked forward to the free time. He published about 20 papers in the10 years after his retirement.

**Acknowledgments:** The work of JPP was supported by the National Science Foundation under grants DMR-2426275 and CHE-2533416. The work of TL was supported by a Feodor Lynen Fellowship from the Humboldt Foundation. The work of RM was supported by the National Science Foundation under grant DMR-2426275. The work of CS was supported by the Department of Energy, Office of Basic Science, under grant no. DE-SCX0018331. The work of AR and JS was supported by the National Science Foundation under grant CHE-2533416. JPP thanks Kieron Burke for pointing out Ref. SG.

Appendix A: Linear Potential Model as a Test of the Gradient Expansion for Exchange

Ref. 39 used the linear potential model of the jellium surface to test the gradient expansion for exchange. In this model, the Kohn-Sham effective potential in the surface region is modelled as

$$v_{eff}(\mathrm{x}) = F(x-a)\theta(x-a), \tag{A1}$$

with

$$F = \frac{1}{2}(\overline{k_F})^2/y_F. \tag{A2}$$

The position $x$=$a$ is chosen to make the electron density neutralize a half-space of uniform positive background for $x < 0$. The electron density tends to the constant $(\overline{k_F})^3/(3\pi^2)$ as $x \to -\infty$, and to 0 as $x \to +\infty$. The density variation in the surface region is rapid when $y_F \to 0$ (infinite barrier mode) and slowly-varying when $y_F \to \infty$. The exact and gradient expansion surface energies $\sigma_x$ are computed and reported in Ref. 39. Table A1 is partly taken from and partly constructed from those values:

Table A1: Values of $\sigma_x/(\overline{k_F})^3 x 1000$ a.u. for the surface exchange energy $\sigma_x$ in the linear potential model of the jellium surface. The "exact" surface exchange energy can have errors of $\pm 1\%$. GEA is the gradient expansion approximation of Eq. (1).

| $y_F$ | $LDA$ | $GEA(C_x^{Sham})$ | $GEA(\left(\frac{10}{7}\right) C_x^{Sham})$ | $Exact$ |
|---|---|---|---|---|
| 0 | 0.894 | -0.316 | -0.835 | 0.576 |
| 1 | 1.252 | 0.880 | 0.692 | 0.899 |
| 2 | 1.609 | 1.293 | 1.158 | 1.268 |
| 4 | 2.439 | 2.219 | . 2.125 | 2.177 |
| 8 | 4.469 | 4.336 | 4.279 | 4.309 |
| 10 | 5.513 | 5.400 | 5.352 | 5.382 |

Table A1 supports the existence of a gradient expansion for exchange, but does not get close enough to the slowly-varying limit to determine whether the correct gradient coefficient is $C_x^{Sham}$ or (10/7) $C_x^{Sham}$. The former is more accurate over the computed range, but only the latter has a monotonic error over this range, and both are accurate at $y_F = 10$.